# Extended Generalized-K (EGK): A New Simple and General Model for Composite Fading Channels

Ferkan Yilmaz and Mohamed-Slim Alouini

Electrical Engineering Program, Division of Physical Sciences and Engineering,

King Abdullah University of Science and Technology (KAUST),

Thuwal, Mekkah Province, Saudi Arabia.

Email(s): {ferkan.yilmaz, slim.alouini}@kaust.edu.sa

## Abstract

In this paper, we introduce a generalized composite fading distribution (termed extended generalized-K (EGK)) to model the envelope and the power of the received signal in millimeter wave (60 GHz or above) and free-space optical channels. We obtain the first and the second-order statistics of the received signal envelope characterized by the EGK composite fading distribution. In particular, expressions for probability density function, cumulative distribution function, level crossing rate and average fade duration, and fractional moments are derived. In addition performance measures such as amount of fading, average bit error probability, outage probability, average capacity, and outage capacity are offered in closed-form. Selected numerical and computer simulation examples validate the accuracy of the presented mathematical analysis.

## I. INTRODUCTION

Radio wave propagation in wireless millimeter wave (60 GHz or above) and free-space optical (FSO) channels is a complicated phenomenon characterized by three interrelated phenomena:

path-loss with distance, shadowing (or long-term fading), and multipath (or short-term) fading. In virtue of fact that shadowing and multipath fading depend on reflection, refraction and scattering while the path-loss is distance dependent [1], [2, references therein], Gaussian-based models are usually employed to model these kind of channels due to their mathematical tractability [1], [2]. Indeed, the Gaussian approximation for the in-phase and quadrature components of the received random vectors leads to the commonly used channel fading models under short-term and long-term fading conditions, for the desired as well as the interfering signals. For instance, short-term fading distribution models include the well-known Rayleigh, Weibull, Rice, Nakagami-m, generalized Nakagami-m models [3]–[6]. On the other hand, long term fading phenomena are modeled by the well-known lognormal distribution [7], [8]. These fading models are typically used to fit the histogram of the empirical/experimental measurements of the envelope of the received random signals. However, there exist fading situations, for which no distribution seems to adequately fit the experimental data, although one or another may yield a moderate fitting. This is accentuated for millimeter wave and FSO channels since multipath (small scale) fading and shadowing (large scale) fluctuations occur simultaneously leading to composite fading channels. In this sense, composite fading models are particularly important for the design of future wireless communications systems [9].

Modeling of composite fading channels is important to comprehend, contemplate and analyze several physical problems in wireless communications including interference effects in cellular systems, multiple input multiple output (MIMO) network, distributed antenna systems, cooperative / multihop relay networks, and optical communications. In a typical signal propagation scenario, the received signal will show fading consisting of very rapid fluctuations around the mean signal level superimposed on relatively slow variations of the mean level. It is therefore not a misstep to assume that the local mean power of the multipath fading is a random variable (RV) distributed over $(0, \infty)$ according to a lognormal or gamma distribution. For example, in $60$ GHz non-line-of-sight (NLOS) propagation, the standard deviation of shadowing is typically larger than that of propagation at $5$ GHz [10], [11] due to the fact that human body (moving objects) shadowing[1] is a significant propagation effect in millimeter wave environment [12], [13]. However, the diffraction effects (propagation of the electromagnetic field behind the obstacle) are significantly smaller in comparison with $5$ GHz band (shadowing zones are very sharp in $60$ GHz) [14]. In short, the environment changes very fast with respect to the diversity and average power of the received signal. In addition, in FSO communication systems, atmospheric turbulence, which originates from variations in the refractive index of the transmission medium due to inhomogeneities in temperature and pressure changes, increases the standard deviation of shadowing on the channel fading the received signal is subjected to. As a result, the identification of a tractable probability density function (PDF) to describe the physical problems at the background of shadowing effects is important in order to study the performance characteristics of emerging and future wireless communication systems operating in this kind of environments.

---

[1]Reported in the literature that signal power can be considerable lost (i.e., up to 20 dB) if the line-of-sight (LOS) component is blocked by a person [12].

In this paper, we focus on a new probability distribution, which is termed extended generalized-K (EGK), to model the fading in wireless millimeter wave channels and FSO environments. The EGK distribution has five parameters and is the extension of the composite fading distribution proposed in [15]–[18]. In addition, the EGK distribution has some good tail properties and includes most of the well-known fading distributions in the literature as either special or limit cases.

The remainder of this paper is organized as follows. In Section II, the EGK distribution is introduced and some of its special cases, which are commonly used in the literature, are outlined. Moreover, the first order statistics of the EGK distribution are derived in closed-forms. Relying upon these first order statistics, Section III contains the second order statistics of the EGK distribution such as level crossing rate (LCR) and average fade duration (AFD). In Section IV, well-known performance measures for digital communications systems such as amount of fading (AoF), average bit error probabilities (ABEP), outage probability (OP), average capacity (AC) and outage capacity (OC) are derived for EGK fading channels, utilizing the results obtained in previous sections. Finally, conclusions are drawn in the last section.

## II. EXTENDED GENERALIZED-K FADING CHANNELS

Let the extended generalized-K (EGK) random process be defined as

$$\mathcal{R}(t) = \underbrace{\left(\left(\frac{\Omega_S}{\beta_s}\right)^{\xi_s} \sum_{\ell=1}^{m_s} \frac{S_\ell^2(t)}{2}\right)^{\frac{1}{2\xi_s}}}_{= S(t)} \underbrace{\left(\left(\frac{\Omega_X}{\beta}\right)^{\xi} \sum_{\ell=1}^{m} \frac{X_\ell^2(t)}{2}\right)^{\frac{1}{2\xi}}}_{= X(t)} \quad (1)$$

where $S_\ell(t)$ and $X_\ell(t)$ are two sets of independent Rayleigh distributed RVs with the average unit powers $\mathbb{E}\left[S_\ell^2(t)\right] = \mathbb{E}\left[X_\ell^2(t)\right] = 1$ such that $S(t)$ and $X(t)$ are a pair of independent RVs representing the shadowing (large-scale) and multipath fading (small-scale) components, respectively. Note that both $S(t)$ and $X(t)$ are distributed according to generalized Nakagami-$m$ PDFs with the average powers $\mathbb{E}\left[S^2(t)\right] = \Omega_S$ and $\mathbb{E}\left[X^2(t)\right] = \Omega_X$, where $\mathbb{E}\left[\cdot\right]$ denotes the expectation operator. In (1), $m$ ($0.5 \leq m < \infty$) and $\xi$ ($0 \leq \xi < \infty$) represent the fading figure (diversity severity / order) and the fading shaping factor, respectively, while $m_s$ ($0.5 \leq m_s < \infty$) and $\xi_s$ ($0 \leq \xi < \infty$) represent the shadowing severity and the shadowing shaping factor (inhomogeneity), respectively. In addition, the parameters $\beta$ and $\beta_s$ are defined as $\beta = \Gamma(m+1/\xi)/\Gamma(m)$ and $\beta_s = \Gamma(m_s+1/\xi_s)/\Gamma(m_s)$, respectively, where $\Gamma(\cdot)$ is the Gamma function [19, Eq. (6.5.3)].

## A. Probability Density Function

The PDF of the received signal envelope $\mathcal{R}$ introduced in (1) is a product of two RVs (that is, $\mathcal{R} = SX$) and as such, using the Mellin transform [20], [21], it is represented in a general compact form given in the following definition.

**Definition 1** (Extended Generalized-K RV). *The distribution $\mathcal{R}$ follows an extended generalized-K (EGK) distribution if the PDF of $\mathcal{R}$ is given by*

$$p_\mathcal{R}(r) = \frac{2\xi}{\Gamma(m_s)\Gamma(m)} \left(\frac{\beta_s\beta}{\Omega}\right)^{m\xi} r^{2m\xi-1} \Gamma\left(m_s - m\frac{\xi}{\xi_s}, 0, \left(\frac{\beta_s\beta}{\Omega}\right)^{m\xi} r^{2\xi}, \frac{\xi}{\xi_s}\right) \quad (2)$$

*where the parameters $m$ ($0.5 \leq m_\ell < \infty$), $\xi$ ($0 \leq \xi_\ell < \infty$), $m_s$ ($0.5 \leq m_\ell < \infty$), $\xi_s$ ($0 \leq \xi_\ell < \infty$) and $\Omega = \Omega_S \Omega_X$ ($0 \leq \Omega < \infty$) are defined above, and where $\Gamma(\cdot,\cdot,\cdot,\cdot)$ is the extended incomplete Gamma function defined as $\Gamma(\alpha, x, b, \beta) = \int_x^\infty r^{\alpha-1} \exp\left(-r - br^{-\beta}\right) dr$, where $\alpha, \beta, b \in \mathbb{C}$ and $x \in \mathbb{R}^+$ [22, Eq. (6.2)].*

In what follows, we utilize a shorthand notation, i.e., $\mathcal{R} \sim \mathcal{K}_\mathcal{G}(m, \xi, m_s, \xi_s, \Omega)$ which denotes that $\mathcal{R}$ follows a EGK distribution with the fading figure $m$, the shaping factor $\xi$, the shadowing figure $m_s$, the shadowing shaping factor $\xi_s$, and the average powers $\Omega$.

To the best of our knowledge, the EGK distribution $\mathcal{R} \sim \mathcal{K}_\mathcal{G}(m, \xi, m_s, \xi_s, \Omega)$ has the advantage of modeling the envelope statistics of most known wireless and optical communication channels. For example, by using [22, Eq. (6.41)] and setting $\xi = 1$ and $\xi = 1$ in $\mathcal{R} \sim \mathcal{K}_\mathcal{G}(m, \xi, m_s, \xi_s, \Omega)$, we get the PDF of generalized K (GK) composite distribution, namely [16]

$$p_{\mathcal{R}}(r) = \frac{4}{\Gamma(m)\Gamma(m_s)} \left(\frac{mm_s}{\Omega}\right)^{\frac{m+m_s}{2}} r^{m+m_s-1} K_{m-m_s}\left(2r\sqrt{\frac{mm_s}{\Omega}}\right). \qquad (3)$$

Moreover, as shadowing severity $m_s$ approaches infinity ($m_s \to \infty$), which means that the distribution of the shadowing $S(t)$ part in (1) follows the dirac's delta distribution ($p_S(r) = \delta(r-\Omega)$), the PDF (2) simplifies into the PDF of the generalized Nakagami-$m$[2], given by [6]

$$p_{\mathcal{R}}(r) = \frac{2\xi}{\Gamma(m)} \left(\frac{\beta}{\Omega}\right)^{m\xi} r^{2m\xi-1} \exp\left(-\left(\frac{m}{\Omega}\right)^{\xi} r^{2\xi}\right), \qquad (4)$$

whose the special or limiting cases are well-known in literature as the Rayleigh ($m_\ell = 1, \xi_\ell = 1$), exponential ($m_\ell = 1, \xi_\ell = 1/2$), Half-Normal ($m_\ell = 1/2, \xi_\ell = 1$), Nakagami-$m$ ($\xi_\ell = 1$), Gamma ($\xi_\ell = 1/2$), Weibull ($m_\ell = 1$), lognormal ($m_\ell \to \infty, \xi_\ell \to 0$), and AWGN ($m_\ell \to \infty, \xi_\ell = 1$). For the other commonly used channel fading models, special or limiting cases of the EGK distribution are listed in Table I. Regarding this great versatility of the EGK distribution, it is important to notice that the EGK distribution offers a kind of unified theory to statistically model the envelope statistics of most known wireless/optical communication channels.

Note that all distributions listed in Table I as the special or limiting cases of the EGK distribution are proposed purely from empirical fitting of measured data to a statistical distribution with their corresponding tail properties $\lim_{r \to \infty} p_{\mathcal{R}}(r)$ and $\lim_{r \to \infty} \partial p_{\mathcal{R}}(r)/\partial r$. If the tail properties approximate to zero for such high $r$ values, i.e., $\lim_{r \to \infty} p_{\mathcal{R}}(r) = 0^+ > 0$ and $\lim_{r \to \infty} \partial p_{\mathcal{R}}(r)/\partial r = 0^- < 0$, then the probability of low amplitude values increase since the condition $\int_0^\infty p_{\mathcal{R}}(r)dr = 1$ is always valid. Explicitly, these tail properties emphasize the distribution at low amplitude values as seen in Fig. 1. For example, the variation of the EGK PDF $p_{\mathcal{R}}(r)$ is depicted in Fig. 1 for 3 dB thresholds (that is, $r = \sqrt{2\Omega}$, $r = \sqrt{\Omega}$ and $r = \sqrt{\Omega/2}$) in order to accentuate how the tails of EGK PDF $p_{\mathcal{R}}(r)$ curves changes with respect not only to both fading figure $m$ and fading shaping factor $\xi$ but also to both fading shaping factor $\xi$ and shadowing shaping factor $\xi_s$. As seen in Fig. 1(a), as either fading figure $m$ or fading shaping factor $\xi$ goes to low possible values ($m \to \frac{1}{2}$ or $\xi \to 0$), the tails of EGK PDF $p_{\mathcal{R}}(r)$ goes to zero, which means that the probability of the distribution at low amplitude values increases since $\int_0^\infty p_{\mathcal{R}}(r)dr = 1$ as mentioned before, and the performance measures of wireless communications systems consequently deteriorates. Additionally, the same consequences are analogously extrapolated for both fading shaping factor $\xi$ and shadowing factor $\xi_s$ as seen in Fig. 1(b).

---

[2]Note that the generalized Nakagami-$m$ RV is the square root of the generalized gamma RV proposed by Stacy [6].

In additive white Gaussian noise (AWGN) channel, the distribution of the instantaneous signal-to-noise ratio (SNR), $\mathcal{G} = \mathcal{R}^2/N_0$ can be directly expressed in terms of average SNR, $\bar{\gamma} \equiv \mathbb{E}[\mathcal{G}] = \mathbb{E}[\mathcal{R}^2]/N_0 = \Omega/N_0$ and $N_0$ representing the power of AWGN noise. The PDF of $\mathcal{G}$, which we term the extended generalized gamma (EGG) PDF, is defined in the following definition.

**Definition 2** (Extended Generalized Gamma RV). *The distribution $\mathcal{G}$ follows an extended generalized gamma (EGG) distribution if the PDF of $\mathcal{G}$ is given by*

$$p_{\mathcal{G}}(\gamma) = \frac{\xi}{\Gamma(m_s)\Gamma(m)} \left(\frac{\beta_s \beta}{\bar{\gamma}}\right)^{m\xi} \gamma^{m\xi-1} \Gamma\left(m_s - m\frac{\xi}{\xi_s}, 0, \left(\frac{\beta_s \beta}{\bar{\gamma}}\right)^{m\xi} \gamma^{\xi}, \frac{\xi}{\xi_s}\right). \quad (5)$$

In what follows, we utilize a shorthand notation, i.e., $\mathcal{G} \sim \mathcal{G}_{\mathcal{G}}(m, \xi, m_s, \xi_s, \bar{\gamma})$ which denotes that $\gamma$ follows a EGG distribution with the fading figure $m$, the shaping factor $\xi$, the shadowing figure $m_s$, the shadowing shaping factor $\xi_s$ and the average powers $\bar{\gamma}$.

## B. Fractional Moments

The fractional moments $\mathbb{E}[\mathcal{R}^k]$, $k \in \mathbb{R}^+$ are crucial for several reasons. First, purely moment-based measures, such as the average SNR and AoF [23], which can be computed using only the first and second central moments of the SNR at the diversity combiner output, are commonly used to characterize the diversity systems. In addition, as shown in [24], higher order moments can also be used to characterize the statistical behavior of the output instantaneous SNR distribution for certain diversity systems. Secondly, more widely used performance measures such as ABEP (which is suitable for digital modulations) and OP, can be typically computed as shown in [25] using the Laguerre moments (computed based on fractional moments) in the case of that the moment generating function (MGF) and PDF of the received instantaneous SNR are not available.

The fractional moments $\mathbb{E}[\mathcal{R}^k]$, $k \in \mathbb{R}^+$ of the signal envelope $\mathcal{R} \sim \mathcal{K}_{\mathcal{G}}(m, \xi, m_s, \xi_s, \Omega)$ are given in the following theorem.

**Theorem 1** (Fractional Moments of EGK RV). *For $k \in \mathbb{R}^+$, the kth moment $\mathbb{E}\left[\mathcal{R}^k\right]$ of the signal envelope $\mathcal{R} \sim \mathcal{K}_\mathcal{G}\left(m, \xi, m_s, \xi_s, \Omega\right)$ is given by*

$$\mathbb{E}\left[\mathcal{R}^k\right] = \frac{\Gamma\left(m_s + \frac{k}{2\xi_s}\right)\Gamma\left(m + \frac{k}{2\xi}\right)}{\Gamma\left(m_s\right)\Gamma\left(m\right)}\left(\frac{\Omega}{\beta_s\beta}\right)^{\frac{k}{2}}. \tag{6}$$

*Proof:* Utilizing [22, Eq. (6.22)] with [26, Eqs. (2.1.4), (2.1.5) and (2.1.11)], the PDF $p_\mathcal{R}(r)$ can also be represented as

$$p_\mathcal{R}(r) = \frac{2}{\Gamma(m_s)\Gamma(m)r}\text{H}_{0,2}^{2,0}\left[\frac{\beta_s\beta}{\Omega}r^2 \middle| \begin{array}{c} --- \\ (m_s, \frac{1}{\xi_s}), (m, \frac{1}{\xi}) \end{array}\right]. \tag{7}$$

where $\text{H}_{p,q}^{m,n}[\cdot]$ is the Fox's H function[3],[4], and where $---$ means that the parameters are absent. Using (7), the kth moment $\mathbb{E}\left[\mathcal{R}^k\right] = \int_0^\infty r^k p_\mathcal{R}(r)\,dr$ can be written as

$$\mathbb{E}\left[\mathcal{R}^k\right] = \frac{2}{\Gamma(m_s)\Gamma(m)}\int_0^\infty r^{k-1}\text{H}_{0,2}^{2,0}\left[\frac{\beta_s\beta}{\Omega}r^2 \middle| \begin{array}{c} --- \\ (m_s, \frac{1}{\xi_s}), (m, \frac{1}{\xi}) \end{array}\right]dr. \tag{8}$$

From the definition of Mellin transform [26, Eq. (2.5.1)] and using [26, Theorem 2.2], the kth moment $\mathbb{E}\left[\mathcal{R}^k\right]$ of the signal envelope $\mathcal{R} \sim \mathcal{K}_\mathcal{G}\left(m, \xi, m_s, \xi_s, \Omega\right)$ is readily obtained as in (6), which proves Theorem 1. ∎

Let us consider some special cases of (6) in order to check its analytical correctness. When setting the fading shaping factor $\xi = 1$ and the shadowing shaping factor $\xi_s = 1$, (6) reduces into [16, Eq. (7)] as expected. For the other commonly used channel fading models, which are listed in Table I as the special or limiting cases of the $\mathcal{R} \sim \mathcal{K}_\mathcal{G}\left(m, \xi, m_s, \xi_s, \Omega\right)$, the kth moment can be readily obtained; for example, when the shadowing severity $m_s$ approaches infinity ($m_s \to \infty$), (6) simplifies into the kth moment of the generalized Nakagami-$m$ RV [29, Eq. (5)].

Using Theorem 1, the kth fractional moment $\mathbb{E}\left[\mathcal{G}^k\right]$, $k \in \mathbb{R}^+$ of $\mathcal{G} \sim \mathcal{G}_\mathcal{G}\left(m, \xi, m_s, \xi_s, \bar{\gamma}\right)$ can be readily obtained as

$$\mathbb{E}\left[\mathcal{G}^k\right] = \mathbb{E}\left[\mathcal{R}^{2k}\right]\bigg|_{\Omega \to \bar{\gamma}} = \frac{\Gamma\left(m_s + \frac{k}{\xi_s}\right)\Gamma\left(m + \frac{k}{\xi}\right)}{\Gamma\left(m_s\right)\Gamma\left(m\right)}\left(\frac{\bar{\gamma}}{\beta_s\beta}\right)^k. \tag{9}$$

---

[3] For more information about the Fox's H function, the readers are referred to [26], [27]

[4] Using [28, Eq. (8.3.22)], the Fox's H function can be represented in terms of the Meijer's G function [28, Eq. (8.2.1)] which is a built-in function in the most popular mathematical software packages such as MATHEMATICA®.

*C. Cumulative Distribution Function*

The CDF $P_\mathcal{R}(r)$ of the signal envelope $\mathcal{R}$ is defined as the probability that the received envelope $\mathcal{R}$ falls below a threshold level $r$ of the signal envelope. In the following theorem, the CDF of $\mathcal{R} \sim \mathcal{K}_\mathcal{G}(m, \xi, m_s, \xi_s, \Omega)$ is given.

**Theorem 2** (Cumulative Distribution Function of EGK RV). *The CDF $P_\mathcal{R}(r)$ of the signal envelope $\mathcal{R} \sim \mathcal{K}_\mathcal{G}(m, \xi, m_s, \xi_s, \Omega)$ is given by*

$$P_\mathcal{R}(r) = \frac{1}{\Gamma(m_s)\Gamma(m)} \mathrm{H}_{1,3}^{3,1}\left[\frac{\beta_s \beta}{\Omega} r^2 \,\bigg|\, \begin{matrix} (1,1) \\ (m_s, \frac{1}{\xi_s}), (m, \frac{1}{\xi}), (0,1) \end{matrix}\right]. \tag{10}$$

*Proof:* Upon having (7), i.e., the PDF of $\mathcal{R} \sim \mathcal{K}_\mathcal{G}(m, \xi, m_s, \xi_s, \Omega)$ in terms of Fox's H function, and utilizing the equality [20, Eq. (4.18)], one can readily obtain the CDF $P_\mathcal{R}(r) = \int_0^r p_\mathcal{R}(r)\,du$ of $\mathcal{R} \sim \mathcal{K}_\mathcal{G}(m, \xi, m_s, \xi_s, \Omega)$ as in (10), which proves Theorem 2. ∎

Note that the PDF of $\mathcal{R} \sim \mathcal{K}_\mathcal{G}(m, \xi, m_s, \xi_s, \Omega)$ given by (10) can also be represented as

$$P_\mathcal{R}(r) = 1 - \frac{1}{\Gamma(m_s)\Gamma(m)} \mathrm{H}_{1,3}^{3,0}\left[\frac{\beta_s \beta}{\Omega} r^2 \,\bigg|\, \begin{matrix} (1,1) \\ (m_s, \frac{1}{\xi_s}), (m, \frac{1}{\xi}), (0,1) \end{matrix}\right]. \tag{11}$$

by means of using the equality given in [20, Eq. (4.17)]. Explicitly, both (10) and (11) are identical functions, but represented differently in terms of Fox's H function. Let us consider special cases of (10); for example, when the shadowing severity $m_s$ approaches infinity ($m_s \to \infty$), we get [29, Eq. (3)] as expected, by mean of substituting $\lim_{a\to\infty} \frac{\Gamma(a+b)a^c}{\Gamma(a+c)a^b} \approx 1$, where $|b| \ll a$ and $|c| \ll a$, into the Mellin-Barnes integral representation [26, Eq. (1.1.1)] of the CDF given by (10) (or (11)).

Note that, in the case of the difficulty in the computation of Fox'H function, (10) can embody the Meijer's G representation for the rational values of the parameters $\xi$ and $\xi_s$ (that is, we let $\xi = k/\ell$ and $\xi_s = k_s/\ell_s$, where $k$, $\ell$, $k_s$ and $\ell_s$ are arbitrary positive integers.) through the medium of algebraic manipulations utilizing [28, Eq. (8.3.2.22)], namely

$$P_\mathcal{R}(r) = \frac{\Phi}{\Gamma(m_s)\Gamma(m)} \mathrm{G}_{k_s k,\, \ell_s k + k_s \ell + k_s k}^{\ell_s k + k_s \ell,\, k_s k}\left[\left(\frac{\beta_s \beta r^2}{\Omega \Psi}\right)^{k_s k} \,\bigg|\, \begin{matrix} -\Xi_{(-k_s k)}^{(k_s k)} \\ \Xi_{(m_s)}^{(\ell_s k)}, \Xi_{(m)}^{(k_s \ell)}, -\Xi_{(1-k_s k)}^{(k_s k)} \end{matrix}\right] \tag{12}$$

with $\Phi = \sqrt{(2\pi)^{2-k\ell_s+k_s\ell}} (k\ell_s)^{m_s-\frac{1}{2}}(k_s\ell)^{m-\frac{1}{2}}$ and $\Psi = (k_s\ell)^{\frac{\ell}{k}}(k\ell_s)^{\frac{\ell_s}{k_s}}$, where $\mathrm{G}_{p,q}^{m,n}[\cdot]$ is the Meijer's G function [28, Eq. (8.3.22)], and the coefficients are defined as $\Xi_{(a)}^{(m)} \equiv \frac{a}{m}, \frac{a+1}{m}, \ldots, \frac{a+m-1}{m}$ with $a \in \mathbb{C}$ and $m \in \mathbb{N}$. Additionally, one can also readily approximate the CDF $P_\mathcal{R}(r)$ with *high* *accuracy* as the sum of PDFs $p_{\mathcal{R}}(r)$ by means of employing GCQ formula [19, Eq. (25.4.39)], which converges rapidly and steadily, requiring few terms for an accurate result. Then, the definite integral in definition of CDF $P_{\mathcal{R}}(r) = \int_0^r p_{\mathcal{R}}(r) du$ can be accurately estimated as

$$P_{\mathcal{R}}(r) = \frac{2\xi \left(\frac{\beta_s \beta}{\Omega}\right)^{m\xi}}{\Gamma(m_s)\Gamma(m)} \sum_{n=1}^{N} \varphi_n \frac{r^{2m\xi}}{\phi_n^{1-2m\xi}} \Gamma\left(m_s - m\frac{\xi}{\xi_s}, 0, \left(\frac{\beta_s \beta}{\Omega}\right)^{m\xi} (r\phi_n)^{2\xi}, \frac{\xi}{\xi_s}\right), \quad (13)$$

where $\varphi_n = \frac{\pi}{2N}\sin\left(\frac{2n-1}{2N}\pi\right)$ and $\phi_n = \frac{1}{2} + \frac{1}{2}\cos\left(\frac{2n-1}{2N}\pi\right)$ for $n \in \{1, 2, \ldots, N\}$, and where the summation stop index $N$ could be chosen as $N = 30$ or more to obtain a high level of accuracy.

Again, using Theorem 2, the CDF $P_{\mathcal{G}}(\gamma)$ of $\mathcal{G} \sim \mathcal{G}_{\mathcal{G}}(m, \xi, m_s, \xi_s, \bar{\gamma})$ can be readily obtained by substituting (10) into $P_{\mathcal{G}}(\gamma) = P_{\mathcal{R}}(\sqrt{\gamma})$ and then changing $\Omega \to \bar{\gamma}$, that is,

$$P_{\mathcal{G}}(\gamma) = \frac{1}{\Gamma(m_s)\Gamma(m)} H_{1,3}^{3,1}\left[\frac{\beta_s \beta}{\bar{\gamma}}\gamma \middle| \begin{array}{c} (1,1) \\ (m_s, \frac{1}{\xi_s}), (m, \frac{1}{\xi}), (0,1) \end{array}\right]. \quad (14)$$

### D. Moment Generating Function

Upon utilizing both alternative exponential forms (i.e., Craig formula forms) of the Gaussian error function, i.e., $\text{erfc}(\sqrt{x}) = (2/\pi)\int_0^{\pi/2} \exp(-x\csc^2(\theta))d\theta$ [5, Eq. (4.2)] and $\text{erfc}^2(\sqrt{x}) = (4/\pi)\int_0^{\pi/4} \exp(-x\csc^2(\theta))d\theta$ [5, Eq. (4.9)], the moment generating function (MGF) of $\mathcal{G} \sim \mathcal{G}_{\mathcal{G}}(m, \xi, m_s, \xi_s, \bar{\gamma})$ is required regarding to the average symbol error probabilities of the receivers operating in wireless communications channels. The MGF of $\mathcal{G} \sim \mathcal{G}_{\mathcal{G}}(m, \xi, m_s, \xi_s, \bar{\gamma})$ is given in closed-form in the following theorem.

**Theorem 3** (Moment Generating Function of EGG RV). *The CDF $\mathcal{M}_{\mathcal{G}}(s)$ of the instantaneous SNR $\mathcal{G} \sim \mathcal{G}_{\mathcal{G}}(m, \xi, m_s, \xi_s, \bar{\gamma})$ is given by*

$$\mathcal{M}_{\mathcal{G}}(s) = \frac{1}{\Gamma(m_s)\Gamma(m)} H_{1,2}^{2,1}\left[\frac{\beta_s \beta}{\bar{\gamma}s} \middle| \begin{array}{c} (1,1) \\ (m_s, \frac{1}{\xi_s}), (m, \frac{1}{\xi}) \end{array}\right]. \quad (15)$$

*Proof:* Upon having (7), i.e., the PDF of $\mathcal{G} \sim \mathcal{G}_{\mathcal{G}}(m, \xi, m_s, \xi_s, \bar{\gamma})$ in terms of Fox's H function, and utilizing the equality [20, Eq. (3.8)], one can readily obtain the MGF $M_{\mathcal{G}}(r) = \int_0^\infty \exp(-s\gamma) p_{\mathcal{G}}(\gamma) d\gamma$ of $\mathcal{G} \sim \mathcal{G}_{\mathcal{G}}(m, \xi, m_s, \xi_s, \bar{\gamma})$ as in (15), which proves Theorem 3. ∎

Note that, upon following the same steps in the derivation of (12), (15) can be represented in terms of Meijer's G function, by favor of [28, Eq. (8.3.2.22)], for the rational values of the shaping parameters $\xi = k/\ell$ and $\xi_s k_s/\ell_s$, where $k$, $\ell$, $k_s$ and $\ell_s$ are arbitrary positive integers, that is,

$$\mathcal{M}_{\mathcal{G}}(s) = \frac{\Phi}{\Gamma(m_s)\Gamma(m)} G_{k_s k,\, \ell_s k + k_s \ell}^{\ell_s k + k_s \ell,\, k_s k} \left[ \left( \frac{\beta_s \beta r^2}{\bar{\gamma}\Psi} \right)^{k_s k} \middle| \begin{array}{c} -\Xi_{(-k_s k)}^{(k_s k)} \\ \Xi_{(m_s)}^{(\ell_s k)}, \Xi_{(m)}^{(k_s \ell)} \end{array} \right]. \tag{16}$$

In addition, setting fading and shadowing shaping parameters to one (i.e., $\xi = 1$ and $\xi_s = 1$) in both (15) and (16), and then using [28, Eq.(8.3.2.21)] produces the MGF of the instantaneous SNR at the output of a receiver operating over generalized-K channels, namely

$$\mathcal{M}_{\mathcal{G}}(s) = \frac{1}{\Gamma(m_s)\Gamma(m)} G_{1,2}^{2,1}\left[ \frac{m_s m}{\bar{\gamma} s} \middle| \begin{array}{c} 1 \\ m_s, m \end{array} \right]. \tag{17}$$

Regarding the special or limiting cases of the EGK distribution listed in Table I, the MGF of the other commonly used channel fading models can be readily found by substituting the corresponding values of the fading figure $m$, the shaping factor $\xi$, the shadowing figure $m_s$, the shadowing shaping factor $\xi_s$ into either (15) or (16).

## III. SECOND ORDER STATISTICS

Referring to (1), when $S(t)$ and $X(t)$ are signal envelopes in some scattering fading channel subjected to Doppler effect in wireless fading channels or to the turbulence effect originate from variations in the refractive index of the transmission channel in free-space optical channels, the signal envelopes $S(t)$ and $X(t)$ are time-correlated random processes and their time derivatives $\dot{S}(t)$ and $\dot{X}(t)$ are not independent from their corresponding envelopes but following zero-mean conditional Gaussian PDFs with respective conditional variances

$$\sigma_{\dot{S}|S}^2 = \frac{\sigma_{\dot{S}_\ell}^2}{2\xi_s^2} \left( \frac{\Omega_S}{\beta_s} \right)^{\xi_s} S^{2\xi_s\left(\frac{1}{\xi_s}-1\right)}, \tag{18a}$$

$$\sigma_{\dot{X}|X}^2 = \frac{\sigma_{\dot{X}_\ell}^2}{2\xi^2} \left( \frac{\Omega_X}{\beta} \right)^{\xi} X^{2\xi\left(\frac{1}{\xi}-1\right)}, \tag{18b}$$

where $\sigma_{\dot{S}_\ell} = \omega \nu_S / \sqrt{2}$ is with the relative vehicle speed $\nu_S$ with respect to the obstructs / keyholes causing shadowing and the wave number given by $\omega = 2\pi/\lambda$ in terms of the wavelength $\lambda$ such that $\omega \nu_S = 2\pi f_S$ where $f_\nu$ is the maximum Doppler frequency shift influencing the shadowing component. Moreover, in (18b), $\sigma_{\dot{X}_\ell} = \omega \nu_X / \sqrt{2}$ is with the relative vehicle speed $\nu_X$ with respect to the transmitter such that $\omega \nu_X = 2\pi f_X$ where $f_X$ is the maximum Doppler frequency shift influencing the multipath fading component. In the following, identical mean power shall be assumed at all components (i.e., $\sigma_{\dot{S}_k}^2 = \sigma_{\dot{S}_\ell}^2$ and $\sigma_{\dot{X}_k}^2 = \sigma_{\dot{X}_\ell}^2$ for $k \neq \ell$). The time derivative of the received envelope $\mathcal{R} \sim \mathcal{K}_{\mathcal{G}}\left(m, \xi, m_s, \xi_s, \Omega\right)$, i.e., $\dot{\mathcal{R}}$ can be obtained by $\dot{\mathcal{R}} = \dot{S}X + S\dot{X} = \dot{S}\mathcal{R}/S + S\dot{X}$. Note that the time derivative $\dot{\mathcal{R}}$ is a zero mean Gaussian RV with the conditional variance $\sigma_{\dot{\mathcal{R}}|RS}^2 = \frac{\mathcal{R}^2}{S^2}\sigma_{\dot{S}|RS}^2 + S^2\sigma_{\dot{X}|RS}^2$, where $\sigma_{\dot{S}|RS}^2 = \sigma_{\dot{S}|S}^2$ and $\sigma_{\dot{X}|RS}^2 = \sigma_{\dot{X}|\frac{\mathcal{R}}{S}}^2 = \sigma_{\dot{X}|X}^2$. For fixed $S = u$ and $\mathcal{R} = r$, the conditional variance is readily obtained as

$$\sigma_{\dot{\mathcal{R}}|RS}^2 = \frac{\sigma_{\dot{S}}^2}{2\xi_s^2}\left(\frac{\Omega_S}{\beta_s}\right)^{\xi_s} u^{2\xi_s(1/\xi_s - 1)} + \frac{\sigma_{\dot{X}}^2}{2\xi^2}\left(\frac{\Omega_X}{\beta}\right)^{\xi}\left(\frac{r}{u}\right)^{2\xi(1/\xi - 1)}. \quad (19)$$

Remembering that the time derivative $\dot{\mathcal{R}}$ is a zero mean Gaussian RV with the conditional variance given in (19), the conditional PDF $p_{\dot{\mathcal{R}}|RS}(\dot{r}|RS)$ can be readily given by

$$p_{\dot{\mathcal{R}}|RS}(\dot{r}|RS) = \frac{1}{\sqrt{2\pi}\sigma_{\dot{\mathcal{R}}|RS}}\exp\left(-\frac{\dot{r}^2}{2\sigma_{\dot{\mathcal{R}}|RS}}\right) \quad (20)$$

distributed over $-\infty < \dot{r} < \infty$.

Finally, utilizing the conditional PDF (20), the second order statistics (i.e., the LCR and the AFD) of the fading envelope $\mathcal{R} \sim \mathcal{K}_{\mathcal{G}}\left(m, \xi, m_s, \xi_s, \Omega\right)$ can be obtained as we show in what follows.

## A. Level Crossing Rate

The LCR of the fading envelope $\mathcal{R} \sim \mathcal{K}_{\mathcal{G}}\left(m, \xi, m_s, \xi_s, \Omega\right)$ at threshold $r$ is defined as the rate at which the EGK random process crosses level $r$ in the negative direction. Using the joint PDF of the EGK envelope random process $\mathcal{R}$ and its time derivative $\dot{\mathcal{R}}$, i.e., $p_{\mathcal{R}\dot{\mathcal{R}}}(r, \dot{r})$, we can obtain the LCR as follows

$$L_{\mathcal{R}}(r) = \int_0^{\infty} \dot{r} p_{\mathcal{R}\dot{\mathcal{R}}}(r, \dot{r}) d\dot{r}, \quad (21)$$

where the joint PDF $p_{\mathcal{R}\dot{\mathcal{R}}}(r, \dot{r})$ can be expressed in terms of conditional PDFs as follows

$$p_{\mathcal{R}\dot{\mathcal{R}}}(r, \dot{r}) = \int_0^{\infty} p_{\dot{\mathcal{R}}|RS}(\dot{r}|r, u) p_{\mathcal{R}|S}(r|u) p_S(u) du, \quad (22)$$

where $p_{\dot{\mathcal{R}}|RS}(\dot{r}|r, u)$ is the conditional Gaussian PDF as mentioned before. Moreover, $p_S(u)$ is the generalized Nakagami-*m* PDF representing the distribution of the shadowing component of the received envelope while $p_{\mathcal{R}|S}(r|u) = p_X(r/u)/u$. Substituting (22) into (21) and using

After some algebraic manipulations, we have

$$L_{\mathcal{R}}(r) = \frac{1}{\sqrt{2\pi}}\int_0^{\infty}\sigma_{\dot{\mathcal{R}}|RS} p_{\mathcal{R}|S}(r|u) p_S(u) du. \quad (23)$$

Substituting (19) into (23), we obtain the exact solution for the LCR as in

$$L_{\mathcal{R}}(r) = 2\sqrt{\frac{2}{\pi} \frac{\xi_s \left(\frac{\beta_s}{\Omega_S}\right)^{\xi_s m_s} \xi \left(\frac{\beta}{\Omega_X}\right)^{\xi m}}{\Gamma(m_s)\Gamma(m)}} \times$$

$$r^{2\xi m-1} \int_0^\infty \sqrt{\frac{\sigma_{\dot{S}_\ell}^2}{2\xi_s^2}\left(\frac{\Omega_S}{\beta_s}\right)^{\xi_s} \frac{r^2}{u^{2\xi_2}} + \frac{\sigma_{\dot{X}_\ell}^2}{2\xi^2}\left(\frac{\Omega_X}{\beta}\right)^{\xi} \frac{u^{2\xi}}{r^{2\xi\left(1-\frac{1}{\xi}\right)}}} \times$$

$$\exp\left(-\left(\frac{\beta_s}{\Omega_S}\right)^{\xi_s} u^{2\xi_s} - \left(\frac{\beta}{\Omega_X}\right)^{\xi}\left(\frac{r}{u}\right)^{2\xi} - 2\left(\xi_s m_s - \xi m\right)\log(u)\right) du. \quad (24)$$

Let us consider some special cases in order to check analytical simplicity and accuracy of our result. As seen in Table I, setting the shadowing and multipath fading shaping factors (inhomogeneity) $\xi_s = 1$ and $\xi = 1$, respectively, in (24) results in the LCR of generalized-K [30, Eq. (9)], as expected. It is useful to mention that the above integral can be computed numerically with desired accuracy (e.g. by using the most popular mathematical software packages such as MATLAB®, MATHEMATICA® and MAPLE™.), or alternatively, it can be readily estimated accurately in virtue of the GCQ rule [19, Eq. (25.4.39)] after changing the variable of the integration in (24) as $u \to \tan(u)$. In addition, the above integral can be approximated by means of the Laplace approximation [30], [31]. Distinctively, we obtained the above integral by utilizing the Taylor series expansion $\sqrt{1+x} = \lim_{N\to\infty} \sum_{n=0}^N \binom{1/2}{n}\left(x^n \theta(1-x) + x^{\frac{1}{2}-n}\theta(x-1)\right)$, where $\theta(\cdot)$ is the Heaviside's theta (unit) function [32, Eq.(1.8.3)]. Shortly, applying the Taylor series expansion of $\sqrt{1+x}$ and performing some algebraic manipulations, (24) can be converted into the sum of two incomplete integrals converging very fast, that is,

$$L_{\mathcal{R}}(r) = \frac{\xi\sqrt{\frac{\Omega_S}{\beta_s}}\sigma_{\dot{S}_\ell}^2 \left(\frac{\beta_s\beta}{\Omega_S\Omega_X}\right)^{m\xi} r^{2m\xi}}{\sqrt{\pi}\xi_s \Gamma(m)\Gamma(m_s)} \lim_{N\to\infty} \sum_{n=0}^N \binom{1/2}{n} \Bigg\{$$

$$\mathcal{Q}_{-n}(r) \int_0^{\mathcal{Q}_{\frac{\xi_s}{\xi_s+\xi}}(r)} u^{(m_s+n)-\frac{\xi}{\xi_s}(m-n)-\frac{\xi_s-1}{2\xi_s}-1} \exp\left(-u - \frac{\sigma_{\dot{X}_\ell}^2 \xi_s^2}{\sigma_{\dot{S}_\ell}^2 \xi^2}\mathcal{Q}_1(r) u^{-\frac{\xi}{\xi_s}}\right) du +$$

$$\mathcal{Q}_{n-\frac{1}{2}}(r) \int_{\mathcal{Q}_{\frac{\xi_s}{\xi_s+\xi}}(r)}^\infty u^{(m_s-n)-\frac{\xi}{\xi_s}(m+n)-\frac{\xi+1}{2\xi_s}-1} \exp\left(-u - \frac{\sigma_{\dot{X}_\ell}^2 \xi_s^2}{\sigma_{\dot{S}_\ell}^2 \xi^2}\mathcal{Q}_1(r) u^{-\frac{\xi}{\xi_s}}\right) du\Bigg\} \quad (25)$$

where the auxiliary function $\mathcal{Q}_n(r)$ is defined as

$$\mathcal{Q}_n(r) = \left(\frac{\sigma_{\dot{S}_\ell}^2 \xi^2}{\sigma_{\dot{X}_\ell}^2 \xi_s^2}\left(\frac{\beta_s\beta}{\Omega_S\Omega_X}\right)^{\xi} r^{2\xi}\right)^n \quad (26)$$

Eventually, substituting [22, Eqs.(6.1) and (6.2)] into (25) results the closed form of the LCR for the fading envelope $\mathcal{R} \sim \mathcal{K}_\mathcal{G}(m, \xi, m_s, \xi_s, \Omega)$ as follows

$$L_\mathcal{R}(r) = \sqrt{\frac{\Omega_S}{\pi \beta_s} \frac{\xi^2}{\xi_s^2} \sigma_{\dot{S}_\ell}^2} \frac{r^{2m\xi} \left(\frac{\beta_s \beta}{\Omega_S \Omega_X}\right)^{m\xi}}{\Gamma(m)\Gamma(m_s)} \lim_{N \to \infty} \sum_{n=0}^{N} \binom{\frac{1}{2}}{n} \Bigg\{$$
$$\mathcal{Q}_{-n}(r) \gamma \left(\mathcal{P}_{-n} - \frac{\xi_s - 1}{2\xi_s}, \mathcal{Q}_{\frac{\xi_s}{\xi_s + \xi}}(r), \frac{\sigma_{\dot{X}_\ell}^2 \xi_s^2}{\sigma_{\dot{S}_\ell}^2 \xi^2} \mathcal{Q}_1(r), \frac{\xi}{\xi_s}\right) +$$
$$\mathcal{Q}_{n - \frac{1}{2}}(r) \Gamma \left(\mathcal{P}_n - \frac{\xi + 1}{2\xi_s}, \mathcal{Q}_{\frac{\xi_s}{\xi_s + \xi}}(r), \frac{\sigma_{\dot{X}_\ell}^2 \xi_s^2}{\sigma_{\dot{S}_\ell}^2 \xi^2} \mathcal{Q}_1(r), \frac{\xi}{\xi_s}\right) \Bigg\} \quad (27)$$

with $\mathcal{P}_n = (m_s - n) - \frac{\xi}{\xi_s}(m+n)$, where $\gamma(\cdot, \cdot, \cdot, \cdot)$ is the extended (lower) incomplete Gamma function defined as $\gamma(\alpha, x, b, \beta) = \int_0^x r^{\alpha - 1} \exp(-r - br^{-\beta}) dr$, where $\alpha, \beta, b \in \mathbb{C}$ and $x \in \mathbb{R}^+$ [22, Eq. (6.1)].

In Fig. 2, the LCR for EGK fading channels is depicted for different parameters. The shadowing increases when the shadowing figure $m_S \to \frac{1}{2}$ and the shadowing factor $\xi_s \to \infty$, or it decreases when the shadowing figure $m_S \to \infty$ and the shadowing factor $\xi_s \to 0$. As seen in Fig. 2, the worst case shadowing is represented by $m_S \to \frac{1}{2}$ and $\xi_s = 1$. The shadowing and multipath fading figures $\{m_s, m\}$ and/or $\{\xi_s, \xi\}$ increase, the LCR decreases as expected.

It may be useful to note that the LCR given in (27) converges rapidly. Indeed, choice $N = 1$ is enough for an approximation. As such, (27) can be readily approximated as

$$L_\mathcal{R}(r) \approx \sqrt{\frac{\Omega_S}{\pi \beta_s} \frac{\xi^2}{\xi_s^2} \sigma_{\dot{S}_\ell}^2} \frac{r^{2m\xi} \left(\frac{\beta_s \beta}{\Omega_S \Omega_X}\right)^{m\xi}}{\Gamma(m)\Gamma(m_s)} \Bigg\{$$
$$\mathcal{Q}_0(r) \gamma \left(\mathcal{P}_0 - \frac{\xi_s - 1}{2\xi_s}, \mathcal{Q}_{\frac{\xi_s}{\xi_s + \xi}}(r), \frac{\sigma_{\dot{X}_\ell}^2 \xi_s^2}{\sigma_{\dot{S}_\ell}^2 \xi^2} \mathcal{Q}_1(r), \frac{\xi}{\xi_s}\right) +$$
$$\mathcal{Q}_{-\frac{1}{2}}(r) \Gamma \left(\mathcal{P}_0 - \frac{\xi + 1}{2\xi_s}, \mathcal{Q}_{\frac{\xi_s}{\xi_s + \xi}}(r), \frac{\sigma_{\dot{X}_\ell}^2 \xi_s^2}{\sigma_{\dot{S}_\ell}^2 \xi^2} \mathcal{Q}(r, 1), \frac{\xi}{\xi_s}\right) +$$
$$\frac{1}{2} \mathcal{Q}_{-1}(r) \gamma \left(\mathcal{P}_{-1} - \frac{\xi_s - 1}{2\xi_s}, \mathcal{Q}_{\frac{\xi_s}{\xi_s + \xi}}(r), \frac{\sigma_{\dot{X}_\ell}^2 \xi_s^2}{\sigma_{\dot{S}_\ell}^2 \xi^2} \mathcal{Q}_1(r), \frac{\xi}{\xi_s}\right) +$$
$$\frac{1}{2} \mathcal{Q}_{\frac{1}{2}}(r) \Gamma \left(\mathcal{P}_1 - \frac{\xi + 1}{2\xi_s}, \mathcal{Q}_{\frac{\xi_s}{\xi_s + \xi}}(r), \frac{\sigma_{\dot{X}_\ell}^2 \xi_s^2}{\sigma_{\dot{S}_\ell}^2 \xi^2} \mathcal{Q}_1(r), \frac{\xi}{\xi_s}\right) \Bigg\}. \quad (28)$$

*B. Average Fade Duration*

The AFD of the received envelope $\mathcal{R} \sim \mathcal{K}_\mathcal{G}(m, \xi, m_s, \xi_s, \Omega)$ at threshold $r$ is defined as the average duration that the EGK random process representing the envelope of the received signal remains below the level $r$ after crossing that level in the downward direction, that is,

$$T(r) = \frac{P_\mathcal{R}(r)}{L_\mathcal{R}(r)}, \tag{29}$$

Finally, substituting (10) and (27) into (29), one can readily obtain the AFD of the received envelope $\mathcal{R} \sim \mathcal{K}_\mathcal{G}(m, \xi, m_s, \xi_s, \Omega)$ at threshold $r$. The AFD of the received envelope $\mathcal{R} \sim \mathcal{K}_\mathcal{G}(m, \xi, m_s, \xi_s, \Omega)$ at threshold $r$ is depicted in Fig. 3 for some scenarios.

## IV. Performance Measures over EGK Fading Channels

In this section, using the expressions obtained in previous sections, the expressions of AoF, ABEP, OP, AC, and OC for digital communications systems operating over EGK fading channels are given in closed-forms, and their analytical simplicity and accuracy are checked by simulations.

*A. Amount of Fading*

The AoF introduced in [23] as a unified performance measure of the severity of fading, is an important measure for the performance of a diversity system since it can be utilized to parameterize the distribution of the signal-to-noise ratio (SNR) of the received signal. Referring to (9), the AoF of $\mathcal{G} \sim \mathcal{G}_\mathcal{G}(m, \xi, m_s, \xi_s, \bar{\gamma})$ can be computed by

$$AoF \equiv \frac{\text{var}[\mathcal{G}]}{\mathbb{E}[\mathcal{G}]^2} = \frac{\mathbb{E}[\mathcal{G}^2]}{\mathbb{E}[\mathcal{G}]^2} - 1 \tag{30}$$

Then, the AoF of $\mathcal{G} \sim \mathcal{G}_\mathcal{G}(m, \xi, m_s, \xi_s, \bar{\gamma})$ is given in the following theorem.

**Theorem 4** (Amount of Fading of EGG RV)**.** *The AoF of $\mathcal{G} \sim \mathcal{G}_\mathcal{G}(m, \xi, m_s, \xi_s, \bar{\gamma})$ is given by*

$$AoF = \frac{\Gamma(m_s)\Gamma\left(m_s + \frac{2}{\xi_s}\right)\Gamma(m)\Gamma\left(m + \frac{2}{\xi}\right)}{\Gamma^2\left(m_s + \frac{1}{\xi_s}\right)\Gamma^2\left(m + \frac{1}{\xi}\right)} - 1. \tag{31}$$

*Proof:* Using (6), the proof is obvious. ∎

As readily seen in (31), when the fading figures $\{m_s, m\}$ and/or $\{\xi_s, \xi\}$ increase, the AoF decreases as expected.

*B. Average Bit Error Probabilities*

The instantaneous bit error probabilities (IBEP), conditioned on the instantaneous SNR $\mathcal{G} \sim \mathcal{G}_{\mathcal{G}}(m, \xi, m_s, \xi_s, \bar{\gamma})$, in an AWGN channel may be written in compact form as [5, Eq. (8.100)]

$$P_E(\mathcal{G}) = \frac{\Gamma(b, a\mathcal{G})}{2\Gamma(b)}, \quad a, b \in \left\{1, \frac{1}{2}\right\}, \tag{32}$$

where $a$ depends on type of modulation scheme ($\frac{1}{2}$ for orthogonal FSK, $1$ for antipodal PSK), $b$ depends on type of detection technique ($\frac{1}{2}$ for coherent, $1$ for non-coherent), and $\Gamma(\cdot, \cdot)$ is moreover complementary incomplete Gamma function [19, Eq. (6.5.3)], respectively. Using the PDF of the instantaneous SNR $\mathcal{G} \sim \mathcal{G}_{\mathcal{G}}(m, \xi, m_s, \xi_s, \bar{\gamma})$, i.e., (5) and utilizing $P_E(\mathcal{G})$, the ABEP in EGK fading channels is given by

$$\overline{P}_E = \int_0^\infty P_E(\gamma) p_{\mathcal{G}}(\gamma) d\gamma. \tag{33}$$

Eventually, substituting (32) and (5) into (34), the ABEP in EGK fading channels is given in closed-form in the following theorem.

**Theorem 5** (Average Bit Error Probabilities in EGK Fading Channels). *Let $\mathcal{G} \sim \mathcal{G}_{\mathcal{G}}(m, \xi, m_s, \xi_s, \bar{\gamma})$ be the instantaneous SNR at the output the receiver operating in wireless EGK fading channels. Then, the ABEP $\overline{P}_E$ of the receiver is given by*

$$\overline{P}_E = \frac{1}{2\Gamma(b)\Gamma(m_s)\Gamma(m)} \mathrm{H}_{2,3}^{2,2}\left[ \frac{\beta_s \beta}{a\bar{\gamma}} \,\bigg|\, \begin{array}{c} (1-b,1)\,(1,1) \\ (m_s, \frac{1}{\xi_s}), (m, \frac{1}{\xi}), (0,1) \end{array} \right]. \tag{34}$$

*for wireless EGK fading channels.*

*Proof:* Substituting (32) and (5) into (34), the ABEP $\overline{P}_E$ can be given as

$$\overline{P}_E = \frac{\xi\left(\frac{\beta_s \beta}{\bar{\gamma}}\right)^{m\xi}}{2\Gamma(b)\Gamma(m_s)\Gamma(m)} \int_0^\infty \gamma^{m\xi-1} \Gamma(b, ar) \Gamma\left(m_s - m\frac{\xi}{\xi_s}, 0, \left(\frac{\beta_s \beta}{\bar{\gamma}}\right)^{m\xi} \gamma^\xi, \frac{\xi}{\xi_s}\right) d\gamma \tag{35}$$

where substituting Meijer's G and Fox's H function representations of both the incomplete gamma $\Gamma(\cdot, \cdot)$ [28, Eq. (8.4.16.2)] and the extended incomplete gamma $\Gamma(\cdot, \cdot, \cdot, \cdot)$ [22, Eq. (6.22)], we get

$$\overline{P}_E = \frac{1}{2\Gamma(b)\Gamma(m_s)\Gamma(m)} \int_0^\infty \frac{1}{\gamma} \mathrm{G}_{1,2}^{2,0}\left[a\gamma \,\bigg|\, \begin{array}{c} 1 \\ 0, b \end{array}\right] \mathrm{H}_{0,2}^{2,0}\left[\frac{\beta_s \beta}{\bar{\gamma}}\gamma \,\bigg|\, \begin{array}{c} --- \\ (m_s, \frac{1}{\xi_s}), (m, \frac{1}{\xi}) \end{array}\right] d\gamma \tag{36}$$

Finally, representing the Meijer's G function in (36) in terms of Fox's H function representation [26, Eq. (2.9.1)] and applying [26, Theorem 2.9] onto (36) results as in (34), which proves Theorem 5. ∎

To check the analytical correctness, let us consider some special cases of (34). Upon setting the fading shaping factor $\xi \to 1$ and the shadowing shaping factor $\xi_s \to 1$ and utilizing [26, Eq. (2.9.1)], (34) is simplified to

$$\overline{P}_E = \frac{1}{2\Gamma(b)\Gamma(m_s)\Gamma(m)} G_{2,3}^{2,2}\left[\frac{m_s m}{a\bar{\gamma}} \middle| \begin{array}{c} 1-b, 1 \\ m_s, m, 0 \end{array}\right]. \tag{37}$$

which is the ABEP $\overline{P}_E$ of the receiver operating in well-known wireless composite fading channel, i.e., wireless generalized-K fading channels.

When the shadowing severity $m_s$ approaches infinity ($m_s \to \infty$) meaning that there is no shadowing effect in the channels, (34) is simplified to the well-known ABEP of binary digital modulation schemes in generalized Gamma fading channels [33, Eq. (10)], namely

$$\overline{P}_E = \frac{1}{2\Gamma(b)\Gamma(m)} H_{2,2}^{1,2}\left[\frac{\beta}{a\bar{\gamma}} \middle| \begin{array}{c} (1-b,1),(1,1) \\ (m,\frac{1}{\xi}),(0,1) \end{array}\right]. \tag{38}$$

Again, in the special case when the fading shaping factor $\xi \to 1$ and the shadowing severity $m_s \to \infty$, (34) becomes the more familiar expression for the ABEP in a flat fading Nakagami-$m$ channel (see [5, Eq. (8.106)], for example), i.e., using [26, Eq. (2.9.1)], [26, Eq. (2.9.16)] and [28, Eq. (6.7)], we get

$$\overline{P}_E = \frac{1}{2\Gamma(b)\Gamma(m)} G_{2,2}^{1,2}\left[\frac{\beta}{a\bar{\gamma}} \middle| \begin{array}{c} 1-b, 1 \\ m, 0 \end{array}\right], \tag{39a}$$

$$= \left(\frac{a\bar{\gamma}}{m+a\bar{\gamma}}\right)^b \frac{\Gamma(m+b)}{2\Gamma(b)\Gamma(m+1)} \left(\frac{m}{m+a\bar{\gamma}}\right)^m {}_2F_1\left[1, m+b; m+1; \frac{m}{m+a\bar{\gamma}}\right], \tag{39b}$$

where ${}_2F_1[\cdot;\cdot;\cdot]$ is the Gaussian hypergeometric function defined in [28, Eq. (7.2.1.1)]. Furthermore, for the special case $m = \xi = 1$, (39a) further simplifies to the well-known result $\overline{P}_E = \frac{1}{2} - \frac{1}{2}\left(\frac{a\bar{\gamma}}{m+a\bar{\gamma}}\right)^b$ [5].

The analytical simplicity and accuracy of (34) are checked by simulations for the noncoherent BFSK (NCFSK)/ differentially coherent BPSK (DPSK) and coherent BPSK/BFSK in Fig. 4. As seen in Fig. 4, when the fading figure increases, then the ABEP decreases since increasing the fading figure decreases the fading severity of the channels. Furthermore, as the shape parameter increases the PDF of the fading becomes skewed around the average power, which implies that the ABEP decreases.

*C. Outage Probability & Outage Capacity*

The OP is defined as the probability that the instantaneous error rate exceeds a specified value or equivalently that the instantaneous SNR $\mathcal{G}$ falls below a certain specified threshold $\gamma_{th}$, i.e.,

$$P_{out} \triangleq \Pr\left(0 \leq \mathcal{G} < \gamma_{th}\right) = \int_0^{\gamma_{th}} p_{\mathcal{G}}\left(\gamma\right) d\gamma \qquad (40)$$

where $p_{\mathcal{G}}(\gamma)$ is the PDF of the instantaneous SNR $\mathcal{G}$ (see (5)). In other words, the OP $P_{out}$ is the CDF of the RV $\mathcal{G}$ evaluated at $\gamma_{th}$. Then, using (14), the OP $P_{out}$ is easily obtained as $P_{out} = P_{\mathcal{G}}(\gamma_{th})$. In Fig. 5, by using $P_{out} = P_{\mathcal{G}}(\gamma_{th})$ as a function for the normalized outage threshold $\gamma_{th}/\bar{\gamma}$, $P_{out}$ is depicted using different fading figure $m$, fading shaping factor $\xi$, shadowing severity $m_s$ and shadowing shaping factor $\xi_s$. Clearly, as seen in Fig. 5, $P_{out}$ decreases as $m$ or $\xi$ increases because increasing $m$ decreases the fading severity and increasing $\xi$ skews the PDF of the fading around the average power $\bar{\gamma}$.

The OC is another important statistical measure to quantify the spectral efficiency in fading channels, which is defined as the probability that the instantaneous capacity, $C_{\mathcal{G}}$ falls below a certain specified threshold $C_{th}$, i.e., $C_{out} = \Pr\left(0 \leq C_{\mathcal{G}} < C_{th}\right)$. With the aid of [34, Eq. (5.1)], the OC $C_{out}$ can be given in terms of (14) as follows $C_{out} = P_{\mathcal{G}}\left(2^{C_{th}/W} - 1\right)$.

*D. Average Capacity*

We consider a signal's transmission of bandwidth $W$ over AWGN channel, then the Shannon capacity is defined as $C_{\mathcal{G}}(\mathcal{G}) \triangleq W \log_2(1 + \mathcal{G})$, where $\mathcal{G} \sim \mathcal{G}_{\mathcal{G}}(m, \xi, m_s, \xi_s, \bar{\gamma})$ is the received instantaneous SNR. Then, the AC $\overline{C}_{\mathcal{G}}$ can be obtained by averaging $C_{\mathcal{G}}(\mathcal{G})$ with the PDF of $\mathcal{G} \sim \mathcal{G}_{\mathcal{G}}(m, 2\xi, m_s, 2\xi_s, \bar{\gamma})$ with $\bar{\gamma} = \mathbb{E}\left[\mathcal{R}^2/N_0\right] = \Omega/N_0$, i.e.,

$$\overline{C}_{\mathcal{G}} = W \int_0^\infty \log_2(1+\gamma) p_{\mathcal{G}}(\gamma) d\gamma. \qquad (41)$$

where $p_{\mathcal{G}}(\gamma)$ is given in (5). With the aid of [22, Eq. (6.22)], utilizing the Fox's H function representation of (5) and substituting [20, Eq. (2.4.3/1)] into (41), we can obtain the AC in EGK fading channels in the closed form as

$$\overline{C}_{\mathcal{G}} = \frac{W}{\log{(2)}\,\Gamma(m)\Gamma(m_s)} \mathrm{H}_{2,4}^{4,1}\left[\frac{\beta_s \beta}{\overline{\gamma}} \,\middle|\, \begin{array}{c} (0,1),(1,1) \\ (m_s, \frac{1}{\xi_s}),(m, \frac{1}{\xi}),(0,1),(0,1) \end{array}\right]. \qquad (42)$$

Note that, using [28, Eq. (8.3.21)], for the shaping factors $\xi_s = \xi = 1$, (42) simplifies to the average capacity of the generalized-K fading channels, namely

$$\overline{C}_{\gamma} = \frac{W/\log{(2)}}{\Gamma(m_s)\,\Gamma(m)} \mathrm{G}_{2,4}^{4,1}\left[\frac{m_s m}{\overline{\gamma}} \,\middle|\, \begin{array}{c} 0,1 \\ m_s, m, 0, 0 \end{array}\right]. \qquad (43)$$

Note that the AC of the generalized-K fading channels given in (43) is in a more compact form than in [18, Eq. (9)]. In addition, when we set the shadowing shaping and figure parameters as $\xi_s = 1 \ m_s \to \infty$, referring to Table I and using [19, Eq. (6.1.46)] in the Mellin-Barnes representation of (42) [26, Eq. (1.1.1)], (42) simplifies into the AC of Nakagami-*m* fading channels given by $\overline{C}_{\gamma} = \frac{W}{\log(2)\Gamma(m)} \mathrm{G}_{2,3}^{3,1}\left[\frac{m}{\overline{\gamma}} \,\middle|\, \begin{array}{c} 0,1 \\ m,0,0 \end{array}\right]$ [35, Eq. (3)]. As seen in Fig. 6, $\overline{C}_{\gamma}$ improves with an increase of fading figures $\{m_s, m\}$ and shaping factors $\{\xi_s, \xi\}$, as expected.

## V. Conclusion

In this paper, we introduced a very general composite fading distribution to model the envelope and power of the received signal in fading channels, which we term the EGK composite fading distribution. We also studied second order of statistics such as level crossing rate and average fade duration for EGK fading channels. We finally obtained closed-form expressions for the amount of fading, average bit error probability, outage probability, outage capacity and average capacity of the EGK fading channels.

## Acknowledgments

This work was supported by King Abdullah University of Science and Technology (KAUST).

[24] M. Z. Win, R. Mallik, and G. Chrisikos, "Higher order statistics of antenna subset diversity," *IEEE Transactions on Wireless Communications*, vol. 2, no. 5, pp. 871–875, Sep. 2003.

[25] F. Yilmaz and M.-S. Alouini, "Sum of Weibull variates and performance of diversity systems," in *Proceeding of the 5th International Wireless Commun. and Mobile Computing Conf. (IWCMC 2009), Leipzig, Germany*, June 21-24 2009.

[26] A. Kilbas and M. Saigo, *H-Transforms: Theory and Applications*.   Boca Raton, FL: CRC Press LLC, 2004.

[27] A. M. Mathai, R. K. Saxena, and H. J. Haubold, *The H-Function: Theory and Applications*, 1st ed.   Dordrecht, Heidelberg, London, New York: Springer Science, 2009.

[28] A. P. Prudnikov, Y. A. Brychkov, and O. I. Marichev, *Integral and Series: Volume 3, More Special Functions*.   CRC Press Inc., 1990.

[29] F. Yilmaz and M.-S. Alouini, "Product of the powers of generalized Nakagami-m variates and performance of cascaded fading channels," in *Proceeding of the IEEE Global Communications Conference (GLOBECOM 2009) - Honolulu, Hawaii*, Nov. 30-Dec. 4 2009.

[30] N. Zlatanov, Z. Hadzi-Velkov, and G. Karagiannidis, "Level crossing rate and average fade duration of the double Nakagami-m random process and application in MIMO keyhole fading channels," *IEEE Communications Letters*, vol. 12, no. 11, pp. 822–824, Nov. 2008.

[31] R. Wong, *Asymptotic Approximations of Integrals*.   SIAM: Society for Industrial and Applied Mathematics, 2001.

[32] D. Zwillinger, *CRC Standard Mathematical Tables and Formulae*, 31st ed.   Boca Raton, FL: Chapman & Hall/CRC, 2003.

[33] V. Aalo, T. Piboongungon, and C.-D. Iskander, "Bit-error rate of binary digital modulation schemes in generalized Gamma fading channels," *IEEE Communications Letters*, vol. 9, no. 2, pp. 139–141, Feb. 2005.

[34] A. Papoulis, *Probability, Random Variables and Stochastic Processes*, 3rd ed.   New York: McGraw-Hill, Feb. 1991.

[35] N. C. Sagias, G. S. Tombras, and G. K. Karagiannidis, "New results for the Shannon channel capacity in generalized fading channels," *IEEE Communications Letters*, vol. 9, no. 2, pp. 97–99, Feb. 2005.

[36] H. Suzuki, "A statistical model for urban radio propogation," *IEEE Transactions on Communications*, vol. 25, no. 7, pp. 673–680, July 1977.

[37] M. H. Ismail and M. M. Matalgah, "Outage probability in multiple access systems with Weibull-faded lognormal-shadowed communication links," in *Proceeding of the IEEE 62nd Vehicular Technology Conference (VTC-2005-Fall) - Dallas, TX*, vol. 4, Sept. 2005, pp. 2129–2133.

[38] P. Bithas, "Weibull-Gamma composite distribution: An alternative multipath/shadowing fading model," *Electronics Letters*, vol. 45, no. 14, pp. 749–751, July 2009.

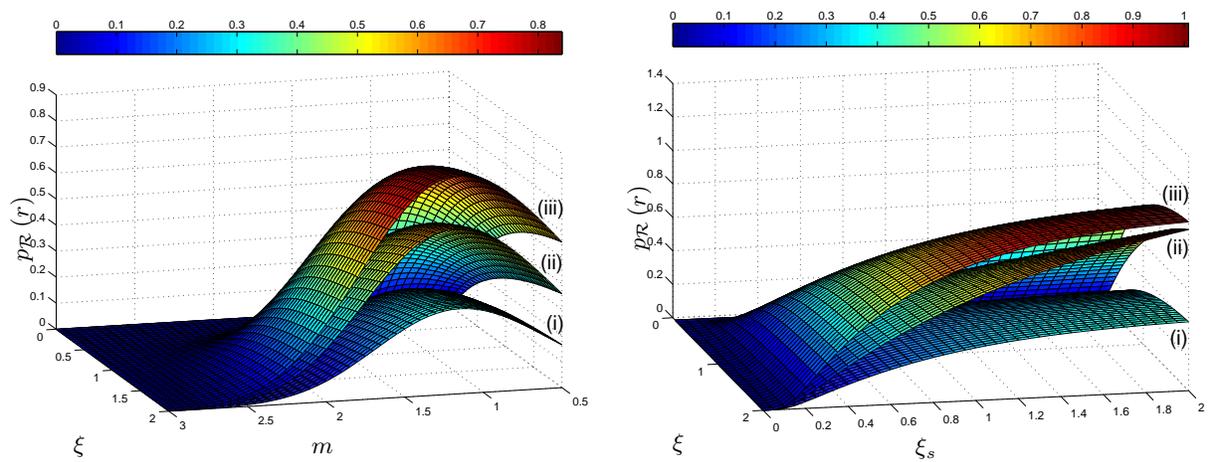

(a) With respect to the fading severity $m$ and the fading shaping factor $\xi$.

(b) With respect to the fading shaping factor $\xi$ and the shadowing shaping factor $\xi_s$.

Fig. 1. The variation of the EGK PDF $p_{\mathcal{R}}(r)$ for $\mathcal{R} \sim \mathcal{K}_{\mathcal{G}}(m, \xi, m_s, \xi_s, \Omega)$ is given with respect not only to both fading figure $m$ and fading shaping factor $\xi$ but also to both fading shaping factor $\xi$ and shadowing shaping factor $\xi_s$, at $3dB$ envelopes: (i) $r = \sqrt{2\Omega}$, (ii) $r = \sqrt{\Omega}$, (iii) $r = \sqrt{\Omega/2}$, where $\Omega = \mathbb{E}[\mathcal{R}^2]$.

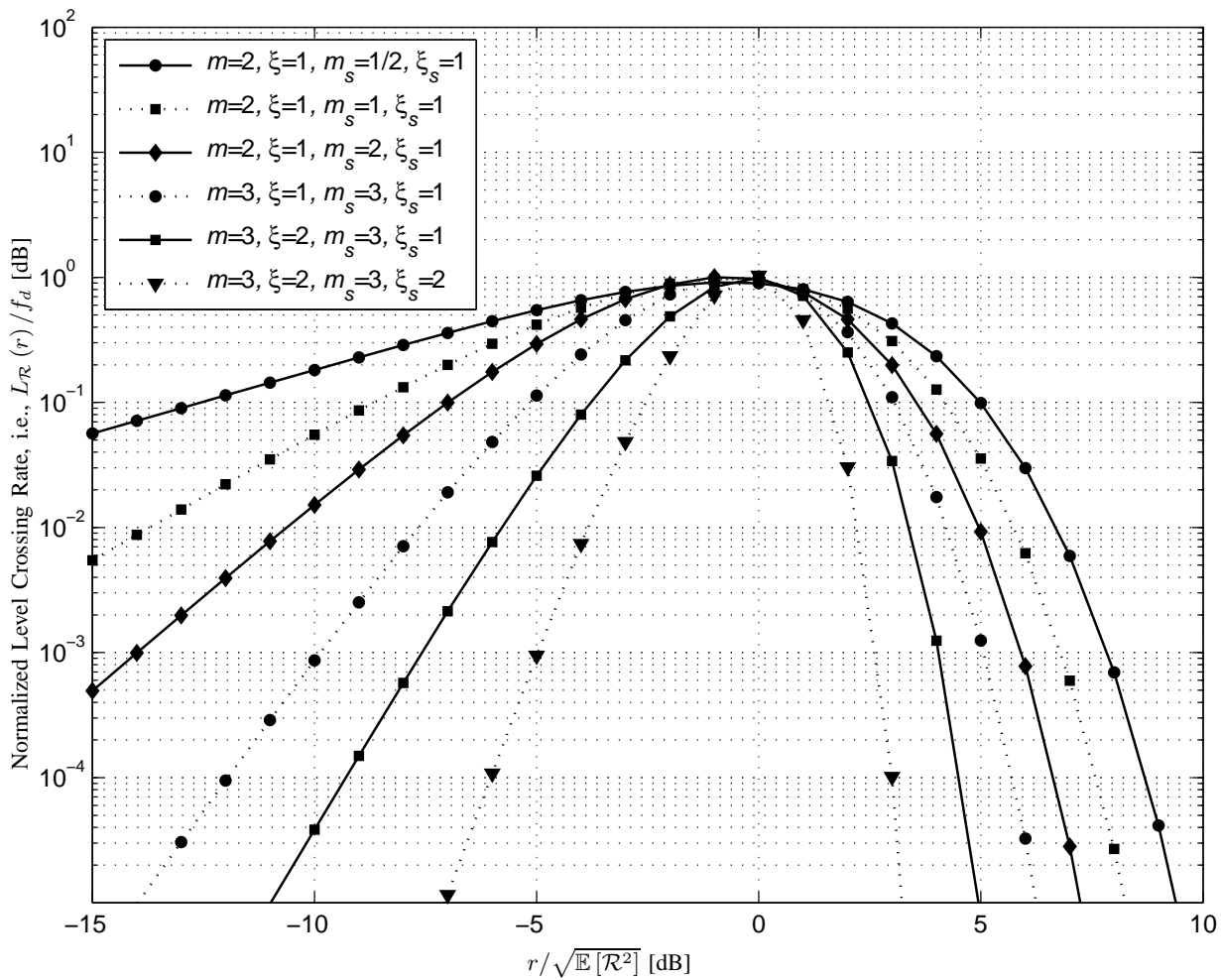

Fig. 2. Normalized LCR versus normalized envelope level in EGK fading channels, where the doppler frequency shifts of multipath and shadowing components of the fading are assumed $f_d = f_S = f_X$.

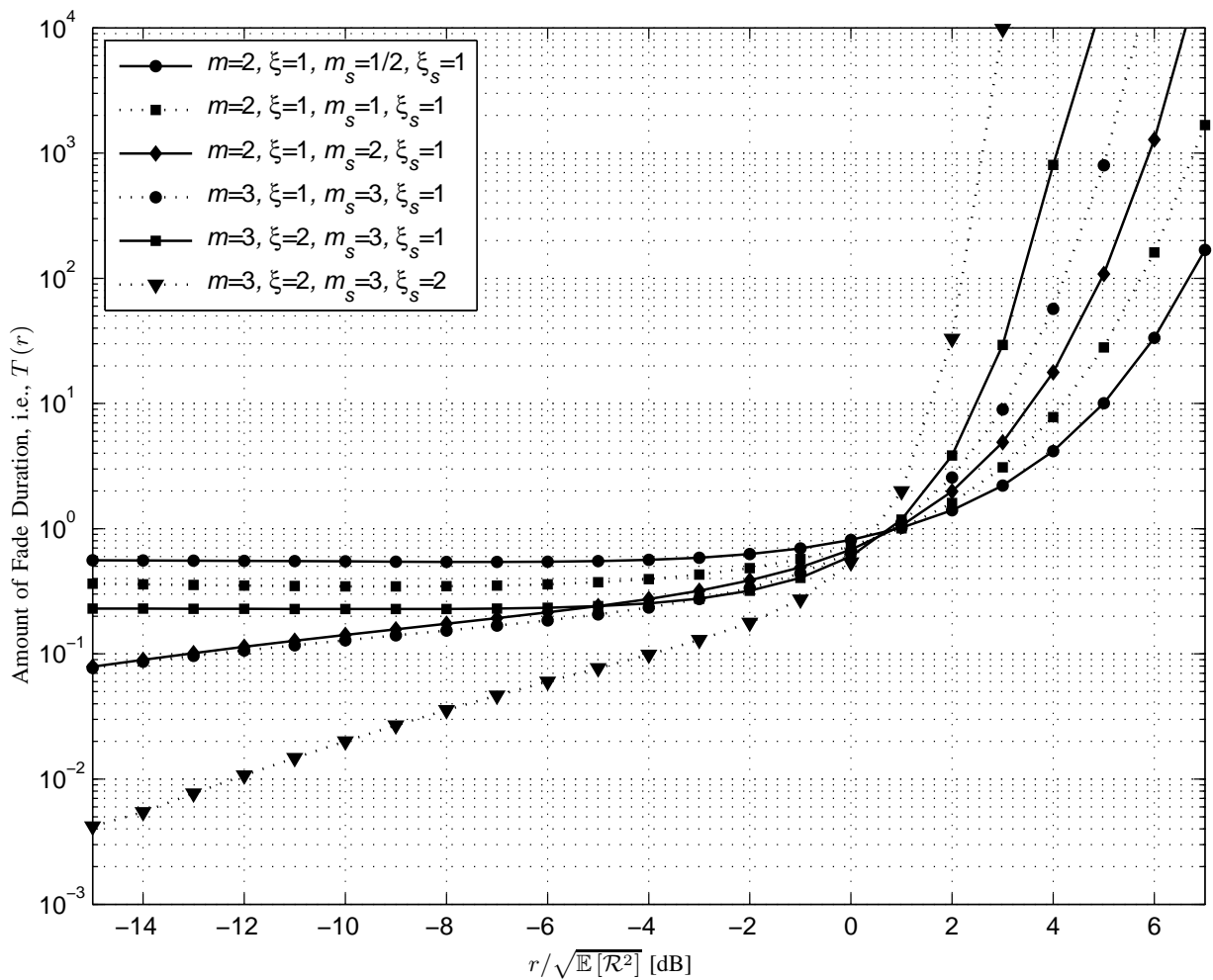

Fig. 3. Normalized AFD versus the normalized envelope level $r/\sqrt{\mathbb{E}\left[\mathcal{R}^2\right]}$ in EGK fading channels, where $r$ is the level and $\mathbb{E}\left[\mathcal{R}^2\right]$ is such a average power as $\mathbb{E}\left[\mathcal{R}^2\right] = \Omega_X \Omega_S$.

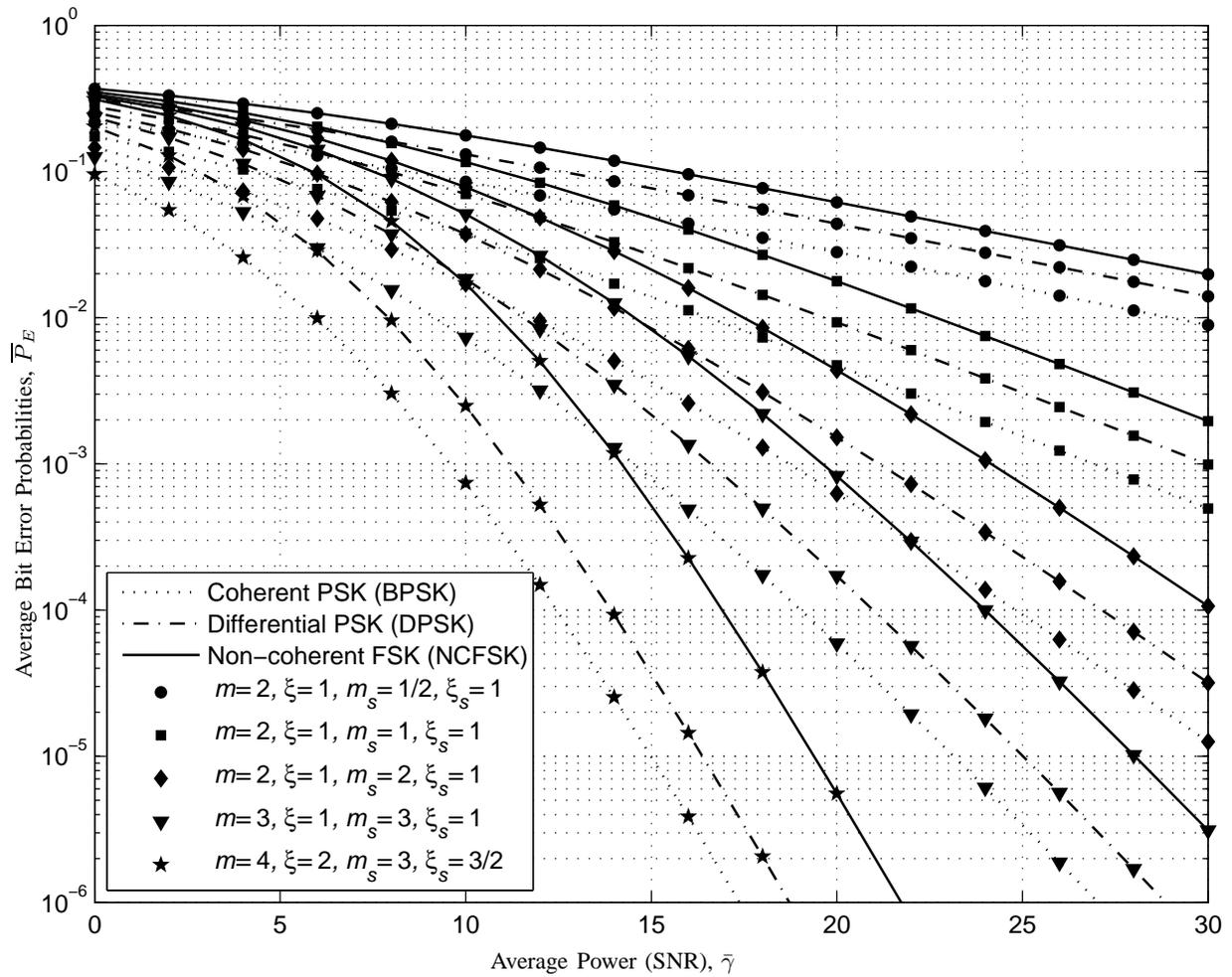

Fig. 4. Average bit error propbabilities of BPSK, DPSK and NCFSK binary modulation schemes over EGK fading channels, i.e., analysis of Eq.(34).

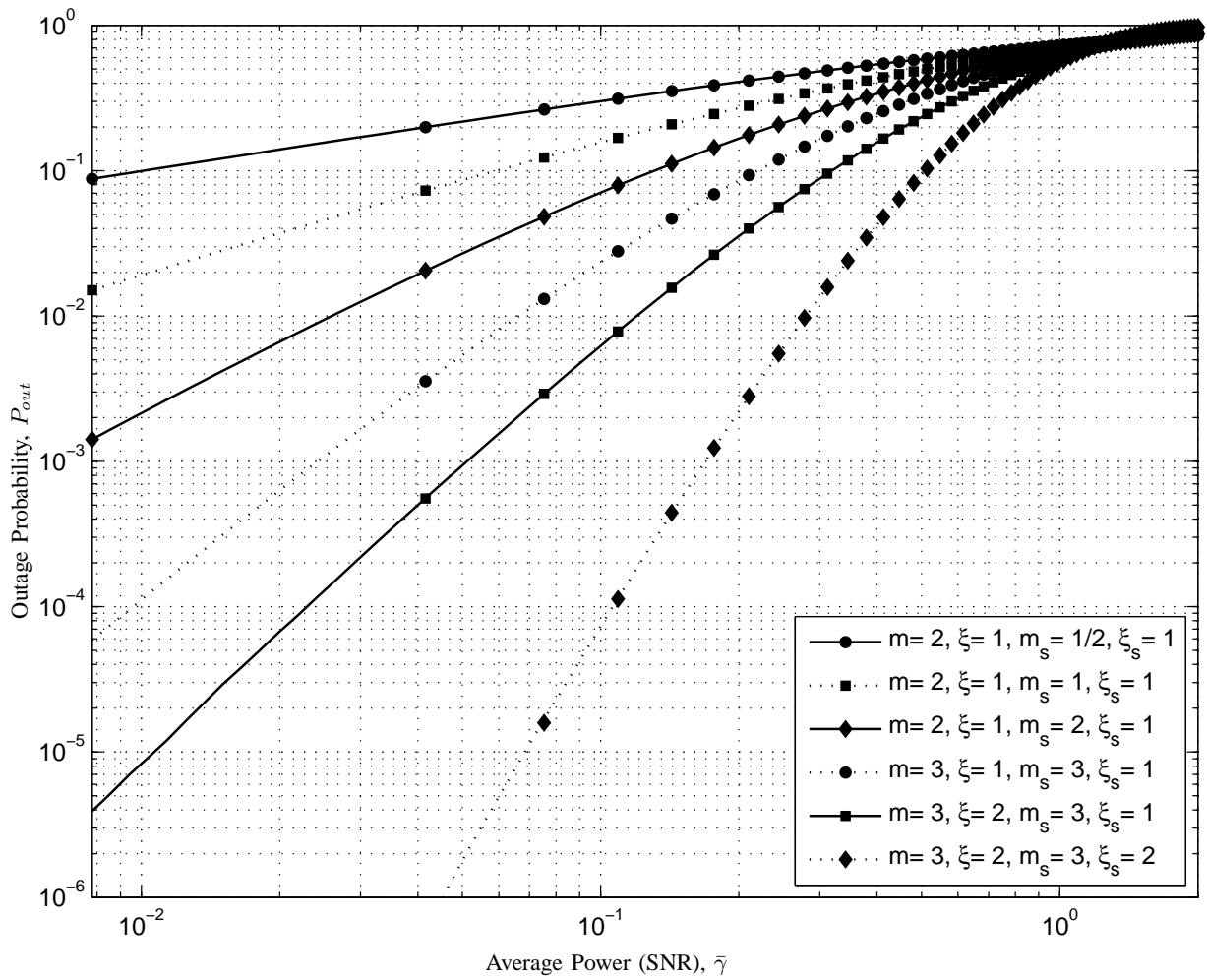

Fig. 5. Outage probability over EGK fading channels, i.e., analysis of Eq.(40) using (14).

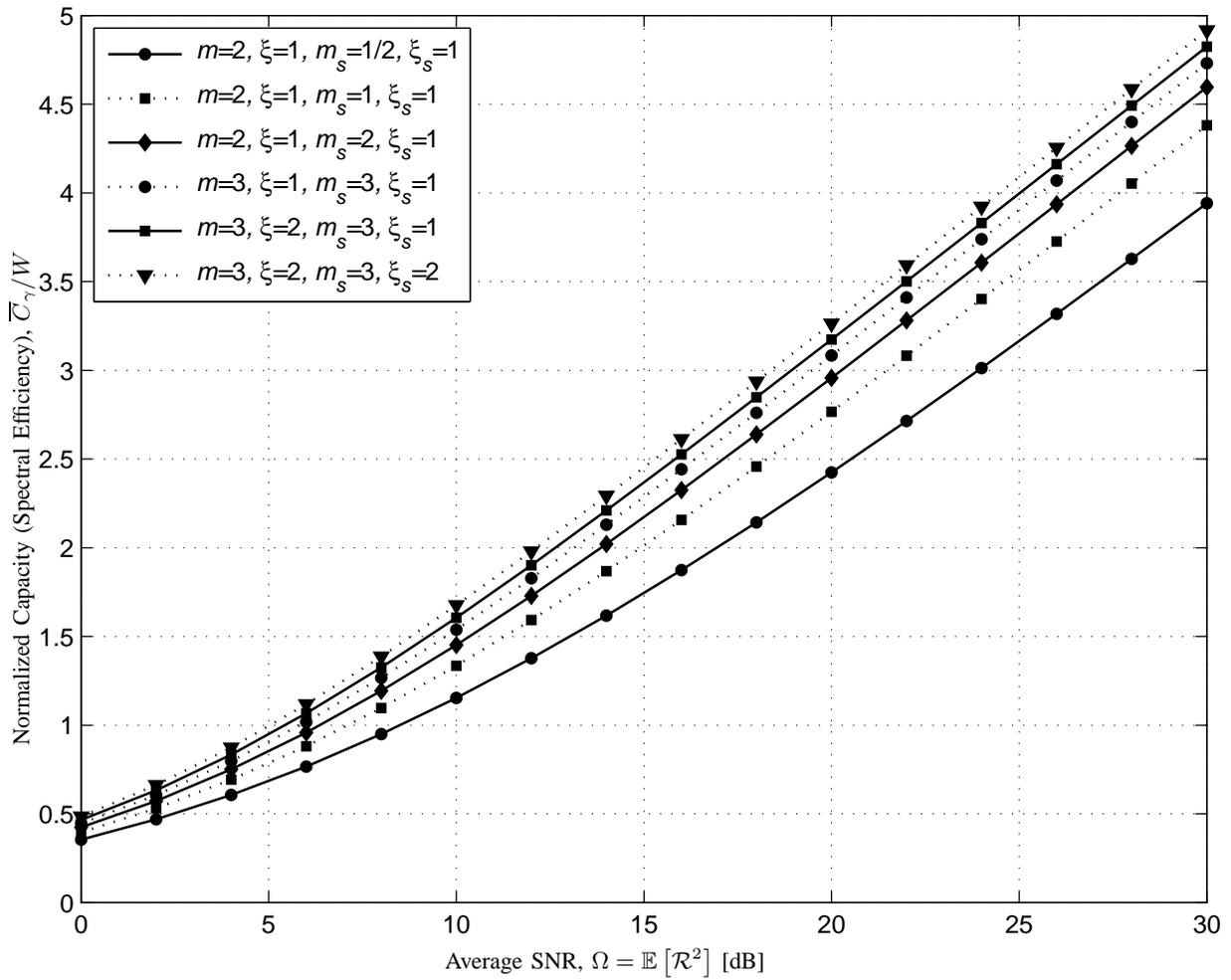

Fig. 6. Normalized capacity $\overline{C}_\gamma/W$ in the EGK fading channels with respect to average SNR $\Omega = \mathbb{E}\left[\mathcal{R}^2\right]$, i.e., analysis of Eq.(42).

TABLE I
SOME SPECIAL CASES OF THE EGK DISTRIBUTION.

| Envelope Distribution | $m$ | $\xi$ | $m_s$ | $\xi_s$ | $\Omega$ |
|---|---|---|---|---|---|
| Rayleigh | 1 | 1 | $\infty$ | 1 | $\Omega$ |
| | $\infty$ | 1 | 1 | 1 | $\Omega$ |
| Maxwell | 3/2 | 1 | $\infty$ | 1 | $\Omega$ |
| | $\infty$ | 1 | 3/2 | 1 | $\Omega$ |
| Half-Normal | 1/2 | 1 | $\infty$ | 1 | $\Omega$ |
| | $\infty$ | 1 | 1/2 | 1 | $\Omega$ |
| Exponential | 1 | 1/2 | $\infty$ | 1 | $\Omega$ |
| | $\infty$ | 1 | 1 | 1/2 | $\Omega$ |
| Weibull | 1 | $\xi$ | $\infty$ | 1 | $\Omega$ |
| | $\infty$ | 1 | 1 | $\xi_s$ | $\Omega$ |
| Nakagami-$m$ | $m$ | 1 | $\infty$ | 1 | $\Omega$ |
| | $\infty$ | 1 | $m_s$ | 1 | $\Omega$ |
| GNM [29] | $m$ | $\xi$ | $\infty$ | 1 | $\Omega$ |
| | $\infty$ | 1 | $m_s$ | $\xi_s$ | $\Omega$ |
| Gamma | $m$ | 1/2 | $\infty$ | 1 | $\Omega$ |
| | $\infty$ | 1 | $m_s$ | 1/2 | $\Omega$ |
| Generalized Gamma [6], [33] | $m$ | $\xi/2$ | $\infty$ | 1 | $\Omega$ |
| | $\infty$ | 1 | $m_s$ | $\xi_s/2$ | $\Omega$ |
| Lognormal | $\infty$ | 0 | $\infty$ | 1 | $\Omega$ |
| | $\infty$ | 1 | $\infty$ | 0 | $\Omega$ |
| Half-Normal-Exponential | 1/2 | 1 | 1 | 1 | $\Omega$ |
| | 1 | 1 | 1/2 | 1 | $\Omega$ |
| Half-Normal-Gamma | 1/2 | 1 | $m_s$ | 1 | $\Omega$ |
| | $m$ | 1 | 1/2 | 1 | $\Omega$ |
| GNM-Lognormal | $\infty$ | 0 | $m_s$ | $\xi_s$ | $\Omega$ |
| | $m$ | $\xi$ | $\infty$ | 0 | $\Omega$ |
| Suzuki [36] | $\infty$ | 0 | 1 | 1 | $\Omega$ |
| | 1 | 1 | $\infty$ | 0 | $\Omega$ |
| Rayleigh-Exponential | 1 | 1 | 1 | 1 | $\Omega$ |
| Maxwell-Lognormal | $\infty$ | 0 | 3/2 | 1 | $\Omega$ |
| | 3/2 | 1 | $\infty$ | 0 | $\Omega$ |
| Maxwell-Exponential | 1 | 1 | 3/2 | 1 | $\Omega$ |
| | 3/2 | 1 | 1 | 1 | $\Omega$ |
| Maxwell-Gamma | $m$ | 1 | 3/2 | 1 | $\Omega$ |
| | 3/2 | 1 | $m_s$ | 1 | $\Omega$ |
| Weibull-Lognormal [37] | $\infty$ | 0 | 1 | $\xi_s$ | $\Omega$ |
| | 1 | $\xi$ | $\infty$ | 0 | $\Omega$ |
| Weibull-Exponential | 1 | 1 | 1 | $\xi_s$ | $\Omega$ |
| | 1 | $\xi$ | 1 | 1 | $\Omega$ |
| Weibull-Weibull | 1 | $\xi$ | 1 | $\xi_s$ | $\Omega$ |
| Weibull-Gamma [38] | $m$ | 1 | 1 | $\xi_s$ | $\Omega$ |
| | 1 | $\xi$ | $m_s$ | 1 | $\Omega$ |
| Nakagami-Lognormal | $\infty$ | 0 | $m_s$ | 1 | $\Omega$ |
| | $m$ | 1 | $\infty$ | 0 | $\Omega$ |
| K-Distribution [15] | $m$ | 1 | 1 | 1 | $\Omega$ |
| | 1 | 1 | $m_s$ | 1 | $\Omega$ |
| Generalized-K [16] | $m$ | 1 | $m_s$ | 1 | $\Omega$ |
| GNM-Exponential | 1 | 1 | $m_s$ | $\xi_s$ | $\Omega$ |
| | $m$ | $\xi$ | 1 | 1 | $\Omega$ |
| GNM-Weibull | $m$ | $\xi$ | 1 | $\xi_s$ | $\Omega$ |
| | 1 | $\xi$ | $m_s$ | $\xi_s$ | $\Omega$ |
| GNM-Gamma | $m$ | 1 | $m_s$ | $\xi_s$ | $\Omega$ |
| | $m$ | $\xi$ | $m_s$ | 1 | $\Omega$ |